\documentclass{article}

\usepackage{PRIMEarxiv}

\usepackage[utf8]{inputenc} 
\usepackage[T1]{fontenc}    
\usepackage[hidelinks]{hyperref} 
\usepackage{url}            
\usepackage{booktabs}       
\usepackage{amsfonts}       
\usepackage{nicefrac}       
\usepackage{microtype}      
\usepackage{lipsum}
\usepackage{fancyhdr} 
\usepackage{amsmath}
\usepackage{amssymb}
\usepackage{graphicx}  

\title{Reinventing the Single-pixel Imaging Paradigm via Quantum-Operator-Based Signal Processing
}

\author{
 Bu-Ran Yu, Yi-Zhu Zhang, Kan-Xu Jia, Qian-Qian Bao, Shao-Ying Meng$^{*}$ and Xi-Hao Chen$^{*}$ \\
  Key Laboratory of Optoelectronic Devices and Detection Technology, School of Physics \\
  Liaoning University \\
  Shenyang 110036, China\\
  \texttt{mengshaoying@163.com; xi-haochen@163.com} \\
}

\begin{document}
\maketitle

\begin{abstract}
A fundamental bottleneck across modern computational imaging and high-dimensional sensing is the conventional decoupled acquisition-reconstruction hierarchy, which subjects high-dimensional spatial sensing to the classical shot-noise limit and intense computational overhead. As a prominent manifestation of this limitation, single-pixel imaging (SPI) suffers severely from this paradigm. We reinvent this paradigm by introducing a quantum-operator-based SPI theoretical framework driven by coherent signal processing. Within this architecture, the spatial inverse problem is analytically mapped into the eigenvalue spectrum of a quantum operator via tailored light-matter interactions. By analytically synthesizing non-linear reconstruction operators via ultra-shallow quantum architectures, we theoretically demonstrate an exponential decay of spatial approximation errors, completely bypassing traditional linear solvers. This operator-space embedding not only shields reconstruction from noise via a strategic error-saturation zone but also bridges the gap from classical shot-noise scaling $\mathcal{O}(1/\sqrt{N_{\text{ph}}})$ to the ultimate Heisenberg limit $\mathcal{O}(1/N_{\text{ph}})$. Crucially, while formulated within SPI, this coherent operator paradigm fundamentally extends to general photon-starved, high-dimensional imaging modalities. This work establishes a universal theoretical blueprint for next-generation quantum-enhanced sensing, shifting the paradigm from iterative optimization to coherent operator-space evolution.
\end{abstract}


\section{Introduction}
Driven by the evolution of computational imaging, the conventional decoupled acquisition-reconstruction hierarchy has long served as the cornerstone of high-dimensional spatial sensing \cite{1,2,3,4}. A prominent manifestation of this paradigm is single-pixel imaging (SPI), which provides a robust solution for imaging through scattering media \cite{5} and across non-visible spectra where high-resolution focal plane arrays are impractical. However, the transition to photon-starved environments—critical for deep-space missions and non-invasive bio-imaging \cite{7778203,Deng_2024,wu2021imaging}—exposes a critical vulnerability: the classical shot-noise limit. Because conventional SPI architectures follow a rigid "measure-then-process" pipeline, reconstruction fidelity is fundamentally tethered to the classical $1/\sqrt{N_{\mathrm{ph}}}$ scaling law \cite{8,9}. This establishes an unyielding barrier where sampling speed and image fidelity become mutually exclusive, a physical bottleneck that resists even the most advanced compressive sensing optimizations. Consequently, a paradigm shift is required, moving beyond iterative digital post-processing toward a coherent, operator-level integration of sensing and reconstruction.

To overcome this bottleneck, quantum-enhanced technologies offer a transformative path. In this work, we propose a fundamental reinvention of the SPI paradigm by leveraging the mathematical rigor of quantum signal processing (QSP) to establish an operator-based theoretical framework. Unlike classical approaches, where the "measure-then-process" sequence leads to irreversible noise accumulation, quantum-operator architectures allow us to coherently encode physical observables into a governed operator space \cite{PRXQuantum.2.040203}. While traditional quantum-enhanced imaging is often constrained by the preparation of complex entangled states and their vulnerability to environmental decoherence \cite{Moreau2019Imaging}, our approach bypasses these hardware burdens through structured light-matter interactions. By synthesizing specific polynomial transformations within shallow quantum architectures, QSP analytically maps the spatial inverse problem to the task of eigenvalue extraction within a controlled Hilbert space \cite{Dong_2021, PRXQuantum.5.020368, s94k-929p}. The core advantage of this methodology lies in its ability to selectively manipulate target features via coherent quantum interference before final measurement collapse. This pre-measurement processing mechanism effectively shields reconstruction from post-detection noise, bridging the gap from classical scaling to the ultimate Heisenberg limit \cite{s94k-929p, cimini2023experimental}.

At the physical interface level, the embedding of spatial sensing into quantum operator spaces is realized via the dispersive coupling between silicon-vacancy (SiV) color centers and nanophotonic crystal cavities \cite{doi:10.1126/science.adu6894, Knaut2024Entanglement, bhaskar2020experimental}. Under a large cavity-atom detuning, the interaction operates in a non-resonant regime, allowing the spatially integrated bucket intensity—even when originating from a spatially incoherent passive light mixture—to induce a deterministic, non-destructive light-shift on the cavity-embedded spin register \cite{8083204, PhysRevLett.123.183602}. At the algorithmic level, this accumulated phase information serves as the Hamiltonian generator within a controlled Hilbert space, where we utilize an eigenvalue extraction scheme inspired by quantum phase estimation \cite{lloyd2014quantum, s94k-929p}. This mechanism enables the measurement accuracy to surpass the standard quantum limit and approach the Heisenberg limit $\mathcal{O}(1/N_{\text{ph}})$, providing a rigorous physical foundation for high-precision computational imaging in photon-sparse regimes \cite{Giovannetti2006Quantum}. In this work, we establish that this theoretical framework not only optimizes reconstruction fidelity but also fundamentally relaxes the stringent requirements for light-source coherence, marking a decisive step toward practical, quantum-resilient computational sensing.

Crucially, the arctangent-based nonlinear reconstruction mapping derived from this coherent evolution exhibits a unique "error saturation" property \cite{pezze2008mach}. This mechanism functions as a physical-layer automatic gain control: by leveraging the geometric characteristics of coherent evolution, the system's sensitivity to noise is suppressed within the flat regions of the nonlinear function, effectively shielding the signal from quantum projection noise and classical readout fluctuations. Our theoretical analysis proves that this framework not only surpasses the standard quantum limit but also introduces a phase-shifting filtering mechanism—a feature fundamentally absent in classical linear sampling systems—allowing the system to asymptotically recover the ultimate Heisenberg scaling $\mathcal{O}(1/N_{\text{ph}})$ under extremely photon-starved regimes. From a practical standpoint, it is remarkable that this analytical exponential convergence is accomplished within an ultra-shallow circuit regime, where the required QSP sequence length is constrained to $5 \le L \le 10$, ensuring robust compatibility with near-term NISQ-era hardware. Crucially, while formulated here within an SPI configuration, the proposed coherent-operator paradigm provides a general pathway toward broader computational imaging modalities facing severe photon-starved constraints. By replacing iterative digital optimization with coherent operator-space evolution, this work establishes a universal blueprint for quantum-enhanced imaging, shifting the focus from digital post-processing toward physics-enabled information extraction at the measurement level.

The remainder of this paper is structured to unfold the theoretical architecture of our quantum-computational imaging framework. Section II establishes the operator formalism for quantum-operator-based SPI, defining the density matrices for spatial modes and the object operator. Section III details the mapping mechanism of bucket signals to qubit registers via dispersive light-matter interactions in SiV-nanophotonic cavity systems. Section IV introduces a quantum state compression protocol utilizing controlled-NOT gates to extract a compact effective density matrix from high-dimensional redundant subspaces. Section V formulates the QSP framework for eigenvalue extraction and the construction of conditional evolution operators. Section VI derives the non-linear reconstruction mapping based on the arctangent function, enabling high-fidelity image retrieval while demonstrating the robust error-suppression characteristics of our proposed operator-based paradigm. Section VIII provides a theoretical feasibility analysis with physical-resource estimates under realistic noise parameters, including a resource-counted experimental feasibility analysis and a resource-counted quantum advantage bound, demonstrating that the proposed framework remains viable within near-term hardware constraints. Finally, Section IX concludes the paper.

\section{Operator Formalism for SPI}

\begin{figure}[tbh]
    \centering
    \includegraphics[width=1.0\linewidth]{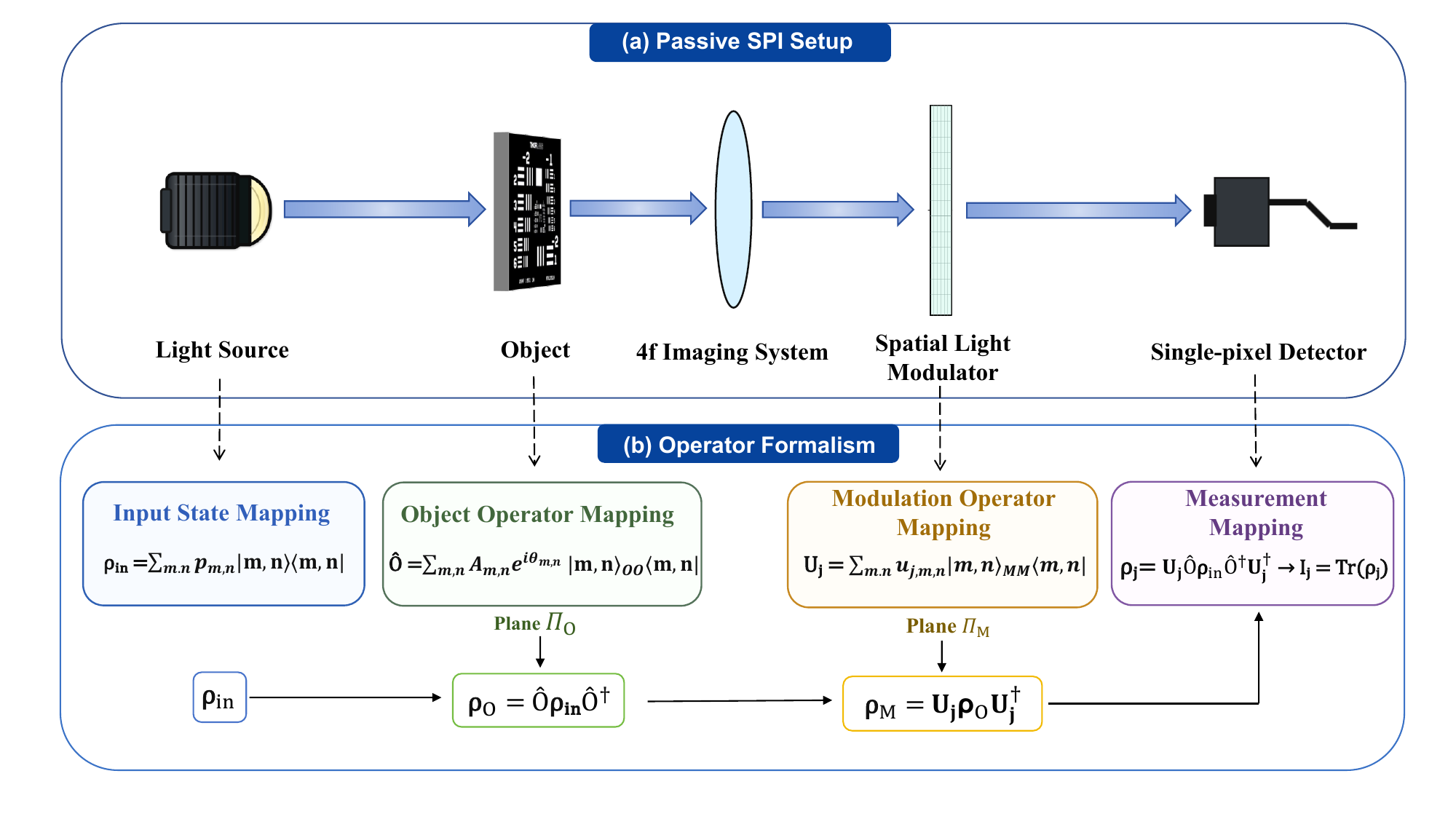}
    \vspace{-1.0cm}  
    \caption{Schematic of (a) passive SPI setup and (b) its corresponding quantum operator formalism.}
    \label{f1}
\end{figure}
As illustrated in Figure.~\ref{f1}(a), we employ passive SPI \cite{Chen_2018,photonics12020164} as the foundational model to investigate the potential of quantum-computational enhancement. The architecture is represented as a sequential optical pipeline: light from the source illuminates the target object, characterized by its spatial distribution $O$. In a discrete representation, the resulting light field is relayed via a 4f system to be mapped directly onto the spatial light modulator (SLM) plane, where it undergoes spatial modulation by a sequence of time-varying patterns $P_i$. The modulated intensity is subsequently integrated and captured by a single-pixel detector \cite{zapata2025design,nano13091542}, yielding a discrete measurement $y_i = \sum_{m,n} O_{m,n} (P_i)_{m,n}$, where $(m,n)$ denotes the spatial mode index. Since each measurement constitutes a mathematical projection of $O$ onto a specific basis $P$, the object is recovered by solving the inverse problem $\hat{O} = \mathcal{R}\{y\}$.

However, within the framework of quantum optics, weak light fields in SPI are no longer treated as classical electromagnetic waves with deterministic intensities; instead, their intrinsic statistical properties must be characterized by quantum states through a rigorous operator formalism, as depicted in Fig.~1(b). We consider an imaging system defined over $N \times N$ discrete spatial modes, where the orthogonal basis vectors are given by the position states $|m,n\rangle$. These basis states correspond to the spatial coordinates $(m,n)$ on the pixelated detection plane, with $m,n \in \{1,\dots,N\}$. In passive SPI, the incident light field is typically spatially incoherent. Consequently, its quantum state is represented as a statistical mixture of various spatial modes, formulated as a density matrix \cite{s94k-929p} given by
\begin{equation}
\rho_{\text{in}} = \sum_{m,n} p_{m,n} |m,n\rangle\langle m,n|,
\end{equation}
where $p_{m,n}$ denotes the probability of a photon occupying the mode $(m,n)$, satisfying the normalization condition $\sum_{m,n} p_{m,n} = 1$. This density matrix is purely diagonal in the position representation, a feature that fundamentally reflects the spatial decoherence of the light field.

To accurately track the state evolution, we explicitly define the spatial bases of the object plane ($\Pi_O$) and the modulation plane ($\Pi_M$) as $\{|m,n\rangle_O\}$ and $\{|m,n\rangle_M\}$, respectively. Upon reaching $\Pi_O$, the light field interacts with the target—a process characterized by the object operator mapping. Specifically, the object operator is represented as
\begin{equation}
\hat{O} = \sum_{m,n} A_{m,n} e^{i\theta_{m,n}} |m,n\rangle_O {}_O\langle m,n|,
\end{equation}
where $A_{m,n} \in [0,1]$ are the real-valued coefficients that, combined with the phase factors $e^{i\theta_{m,n}}$, characterize the complex transmission of the object at mode $(m,n)$. After passing through $\Pi_O$, the quantum state transforms into the density matrix $\rho_O = \hat{O} \rho_{\text{in}} \hat{O}^\dagger$. 

Subsequently, the 4f imaging system relays the field, mapping the spatial bases from $\Pi_O$ to $\Pi_M$. For an ideal mapping, the bases in both planes are mathematically equivalent. During each sampling cycle $j$ ($j = 1, 2, \dots, M$), the SLM imposes a predetermined modulation governed by the modulation operator of
\begin{equation}
U_j = \sum_{m,n} u_{j,m,n} |m,n\rangle_M {}_M\langle m,n|,
\end{equation}
where the complex coefficients $u_{j,m,n}$ satisfy $|u_{j,m,n}| \leq 1$. Following the modulation at $\Pi_M$, the quantum state further evolves into $\rho_M = U_j \rho_O U_j^\dagger$. The total output density matrix corresponding to the $j$-th measurement is thus given by
\begin{equation}
\rho_j = U_j \hat{O} \rho_{\text{in}} \hat{O}^\dagger U_j^\dagger.
\end{equation}
Crucially, since $\hat{O}$, $U_j$, and the initial $\rho_{\text{in}}$ are all diagonal in the position representation, they mutually commute. This property ensures that $\rho_j$ preserves its diagonal structure, reflecting the maintained spatial decoherence throughout the imaging pipeline.

In classical SPI, the detector records the total intensity, which corresponds to the trace of the density matrix: $I_j = \mathrm{Tr}(\rho_j)$. In a quantum-computational-enhanced framework, however, we do not perform a direct classical readout. Instead, our objective is to encode $\rho_j$ as a quantum state into the registers of a quantum processor for further coherent processing.

\section{Mapping of Bucket Signals to Pixel Qubit Registers}
To bridge the gap between classical intensity $I_j = \mathrm{Tr}(\rho_j)$ and quantum registers, we employ a quantum interface based on the dispersive coupling between a  SiV color center and a nanophotonic crystal cavity \cite{bhaskar2020experimental,doi:10.1126/science.adu6894,Knaut2024Entanglement}. In the dispersive regime where the cavity-atom detuning $\Delta$ far exceeds the natural linewidth $\gamma$, the interaction is governed by the effective Hamiltonian
\begin{equation}
H_{\text{eff}} = \hbar \frac{g^2}{\Delta} a^\dagger a \sigma_z,
\end{equation}
where $g$ denotes the single-photon Rabi frequency. This interaction induces a non-destructive light-shift, accumulating a relative phase $\phi_j$ in the SiV spin state proportional to the integrated intensity
\begin{equation}
\phi_j = \frac{g^2 t}{\Delta} \mathrm{Tr}(\rho_j) \equiv k \mathrm{Tr}(\rho_j),
\end{equation}
with $t$ being the interaction time and $k$ a tunable scaling constant.

This phase information is encoded into Register A via a standard interferometric sequence. Starting from the ground state $|0\rangle$, a Hadamard gate prepares the superposition $(|0\rangle + |1\rangle)/\sqrt{2}$. The subsequent phase-encoding gate $U(\phi_j) = \mathrm{diag}(1, e^{i\phi_j})$ maps the intensity onto the relative phase, which is then converted into a population distribution by a second Hadamard gate, yielding the state $|\psi_2\rangle = \frac{1}{2}[(1+e^{i\phi_j})|0\rangle + (1-e^{i\phi_j})|1\rangle]$. The corresponding density matrix $\rho_A = |\psi_2\rangle\langle\psi_2|$ can be expressed in terms of observable populations as
\begin{equation}
\rho_A = 
\begin{pmatrix}
\cos^2\frac{\phi_j}{2} & \frac{i}{2}\sin\phi_j \\
-\frac{i}{2}\sin\phi_j & \sin^2\frac{\phi_j}{2}
\end{pmatrix}, 
\end{equation}
where $\phi_j = k \mathrm{Tr}(\rho_j)$.

To preserve the spatial characteristics of the system, we introduce Register B to encode the modulation pattern $U_j$. By splitting the optical path, the spatial complex amplitude is mapped onto a coherent superposition state $|\psi_{U_j}\rangle = G_j^{-1/2} \sum_k U_j(k) |k\rangle$. Since Registers A and B are physically distinguishable, the composite system is described by the tensor product $\rho_{\text{total}} = \rho_A \otimes \rho_B$, or explicitly:
\begin{equation}
\rho_{\text{total}} = 
\begin{pmatrix}
\cos^2\left(\frac{k\mathrm{Tr}(\rho_j)}{2}\right) \rho_B & \frac{i}{2}\sin\left(k\mathrm{Tr}(\rho_j)\right) \rho_B \\
-\frac{i}{2}\sin\left(k\mathrm{Tr}(\rho_j)\right) \rho_B & \sin^2\left(\frac{k\mathrm{Tr}(\rho_j)}{2}\right) \rho_B
\end{pmatrix}.
\end{equation}

\section{Quantum State Compression}
To facilitate efficient processing, we introduce a memory-assisted auxiliary qubit initially in the ground state $|0\rangle$, expanding the global composite system to $\rho_1 = \rho_{\text{total}} \otimes |0\rangle\langle 0|$. For clarity in tracking the subsequent evolution, we define the block matrix elements based on Eq.~(8) as
\begin{equation}
\begin{aligned}
A &= \cos^2\tfrac{\phi_j}{2} \cdot \rho_B, \quad B = \tfrac{i}{2}\sin\phi_j \cdot \rho_B, \\
C &= -\tfrac{i}{2}\sin\phi_j \cdot \rho_B, \quad D = \sin^2\tfrac{\phi_j}{2} \cdot \rho_B,
\end{aligned}
\end{equation}
where $\phi_j = k \mathrm{Tr}(\rho_j)$. In the basis of Register A and the auxiliary qubit $\{|00\rangle, |01\rangle, |10\rangle, |11\rangle\}$, the initial state $\rho_1$ is a sparse block matrix with non-zero components only in the entries involving the auxiliary $|0\rangle$ state.

We then apply a controlled-NOT (CNOT) operation, $U_{CX} = |0\rangle\langle 0| \otimes I + |1\rangle\langle 1| \otimes X$, where Register A acts as the control and the auxiliary qubit as the target. This gate reconfigures the global coherence by mapping the state $|1, \rho_B, 0\rangle$ to $|1, \rho_B, 1\rangle$. Under the conjugate action $\rho_2 = U_{CX} \rho_1 U_{CX}^\dagger$, the block matrix is rearranged into
\begin{equation}
\rho_2 = 
\begin{pmatrix}
A & 0 & 0 & B \\
0 & 0 & 0 & 0 \\
0 & 0 & 0 & 0 \\
C & 0 & 0 & D
\end{pmatrix}.
\end{equation}
In this representation, the information-carrying elements $\{A, B, C, D\}$ are shifted to the four corners of the global density matrix, corresponding to the subspace spanned by $\{|00\rangle, |11\rangle\}$.
\begin{figure}[t]
    \centering
    \includegraphics[width=0.88\linewidth]{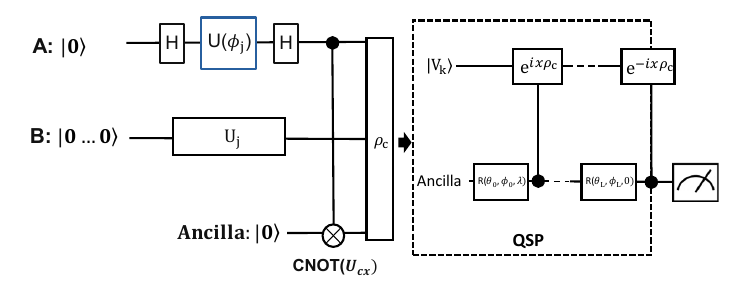}
    \vspace{-0.6cm}
    \caption{Quantum circuit schematic of the proposed pre-measurement processing and QSP readout protocol. Register A is initialized in $|0\rangle$ and undergoes the $H$-$U(\phi_j)$-$H$ interferometric sequence, where the phase gate $U(\phi_j)$ encodes the bucket signal through $\phi_j=kI_j$ with $I_j=\mathrm{Tr}(\rho_j)$. Register B is initialized in $|0\cdots0\rangle$ and loaded with the modulation-pattern information through $U_j$. The auxiliary qubit is prepared in $|0\rangle$ and participates in the CNOT-assisted compression operation $U_{CX}$, which rearranges the information-carrying blocks of the joint state into the compact effective density matrix $\rho_c$. The QSP module then acts on the eigenstate $|V_k\rangle$ and an ancilla qubit by alternating controlled evolutions generated by $\rho_c$, $e^{\pm i x\rho_c}$, with programmable single-qubit rotations $R(\theta_0,\phi_0,\lambda)$ and $R(\theta_L,\phi_L,0)$. The final projective measurement of the ancilla provides the spectral information required for nonlinear intensity reconstruction.}
    \label{f2}
\end{figure}
Since the rows and columns associated with the basis states $|01\rangle$ and $|10\rangle$ remain entirely null and decoupled from the evolution, the system can be isomorphically projected onto a reduced $2 \times 2$ block-dimensional space. This projection yields the compressed effective density matrix
\begin{equation}
\rho_c = \begin{pmatrix} A & B \\ C & D \end{pmatrix} = 
\begin{pmatrix}
\cos^2\frac{\phi_j}{2} \cdot \rho_B & \frac{i}{2}\sin\phi_j \cdot \rho_B \\
-\frac{i}{2}\sin\phi_j \cdot \rho_B & \sin^2\frac{\phi_j}{2} \cdot \rho_B
\end{pmatrix}.
\end{equation}
By distilling the key coherent information into the lowest possible dimension, $\rho_c$ serves as the minimal operational object for subsequent quantum eigenvalue extraction and non-linear image reconstruction.

\section{Quantum Feature Extraction via QSP}
Following the effective compression of the state space, the high-dimensional redundant information is successfully stripped away, encapsulating the core image features within a low-dimensional effective density matrix $\rho_c$. This section elucidates the mechanism by which the intrinsic information related to the bucket signal and speckle patterns is isolated from this density matrix using quantum principal component analysis (QPCA) in conjunction with the QSP framework.

The compressed density matrix $\rho_c$, representing the subspace of the target system, is parameterized in the $2 \times 2$ Hermitian form
\begin{equation}
\rho_c = \begin{pmatrix} A & iE \\ -iE & D \end{pmatrix},
\end{equation}
where the matrix elements are defined by the modulation intensity and the speckle projection operator as
\begin{equation}
A = \cos^2 \tfrac{\phi_j}{2} \cdot \rho_{B}, \quad D = \sin^2 \tfrac{\phi_j}{2} \cdot \rho_{B}, \quad E = \tfrac{1}{2} \sin \phi_j \cdot \rho_{B}.
\end{equation}
Determining the spectral properties of $\rho_c$ via the characteristic equation $\det(\rho_c - rI) = 0$ yields the two eigenvalues $r_{1,2} = (A+D \pm \sqrt{(A-D)^2 + 4E^2})/2$, satisfying the trace normalization $r_1 + r_2 = 1$. The corresponding orthonormal eigenvectors $|V_k\rangle$, associated with each eigenvalue $r_k$, form the invariant basis for the subsequent coherent evolution.

In the QSP framework, $\rho_c$ serves as the Hamiltonian generator to construct the unitary evolution operator $e^{\pm i x \rho_c}$. By expanding the density matrix into the Pauli basis $\{I, \sigma_x, \sigma_y, \sigma_z\}$, the evolution operator is explicitly solved as
\begin{equation}
e^{\pm i x \rho_c} = e^{\pm i \frac{A+D}{2} x}
\begin{pmatrix}
\cos(wx) \pm i \frac{A-D}{2w} \sin(wx) & \mp \frac{E}{w} \sin(wx) \\
\mp \frac{E}{w} \sin(wx) & \cos(wx) \mp i \frac{A-D}{2w} \sin(wx)
\end{pmatrix},
\end{equation}
where $w = \sqrt{(\frac{A-D}{2})^2 + E^2}$. This operation is further upgraded to a controlled unitary gate by introducing an auxiliary control channel, where the control qubit state dictates the direction of time evolution.

To distinguish between the two eigenstates $|V_1\rangle$ and $|V_2\rangle$, we utilize QSP\cite{s94k-929p} to encode the eigenvalue information into the relative phase of an ancilla qubit. By alternating single-qubit rotations on the ancilla with the controlled evolution $e^{\pm i x \rho_c}$, we construct a global unitary $U_{\text{QSP}}$ that implements a polynomial functional mapping
\begin{equation}
U_{\text{QSP}} \left( |V_k\rangle \otimes |0\rangle \right) \approx |V_k\rangle \otimes \left( f_s(r_k x) |0\rangle + g_s(r_k x) |1\rangle \right),
\end{equation}

The phases used in this QSP sequence are obtained from a recursive Fourier-coefficient deflation procedure. 
The target response $f_s(\tau)$ and its complementary polynomial $g_s(\tau)$ are first expanded as finite Laurent series,where $\tau \equiv r_k x$ is the mapped scalar argument.
$f(e^{i\tau})=\sum_{n=0}^{L}a_ne^{in\tau}$ and
$g(e^{i\tau})=\sum_{n=0}^{L}b_ne^{in\tau}$, with the initial coefficients chosen from the Fourier representation of the shifted step filter. 
At recursion level $l$, the leading coefficients determine the programmable rotation angles through
$\theta_l=\arctan(|b_l|/|a_l|)$ and $\phi_l=\arg(a_l/b_l)$.
The highest-order component is then removed by coefficient deflation, reducing the polynomial degree from $l$ to $l-1$.
Iterating this process from $l=L$ to $1$ yields the phase sequence $\{(\theta_l,\phi_l)\}_{l=1}^{L}$, while the remaining zeroth-order coefficient fixes the terminal phase $\lambda=\arg(b_0)$.
Here $f_s(r_kx)$ is chosen to approximate a shifted periodic Heaviside step function.

To ensure unambiguous separation of the eigenstates, we impose the forbidden region condition $r_2 x < s - \Delta < s + \Delta < r_1 x$, where $\Delta$ represents the transition width of the polynomial approximation. This strategic parameterization ensures that the eigenvalues are mapped to opposite sides of the step function, resulting in the non-linear response $f_s(r_1x) \approx 1$ and $f_s(r_2x) \approx 0$. Consequently, the QSP sequence induces a macroscopic distinction between the states
\begin{equation}
U_{\text{QSP}}(|V_1\rangle \otimes |0\rangle) \approx |V_1\rangle \otimes |0\rangle, \quad U_{\text{QSP}}(|V_2\rangle \otimes |0\rangle) \approx |V_2\rangle \otimes |1\rangle.
\end{equation}
The final joint system state $\rho_{\text{3,out}} \approx \sum_k r_k |V_k\rangle\langle V_k| \otimes |k\rangle\langle k|$ exhibits a high degree of entanglement, allowing the deterministic extraction of the target feature associated with $r_2$ via projective measurement on the ancilla qubit.

\section{Image Reconstruction}
\noindent
Having completed the non-linear mapping of the feature space via QSP, this section will explain in detail how to use the measured results of the processed auxiliary bits to map back to the physical space, thereby achieving high-fidelity reconstruction of the target object’s image.

By measuring the auxiliary bits, we obtained the collapsed state associated with the eigenvalue $r_1, r_2$.When the measurement result is $|1\rangle$ , the target system collapses with probability $r_2$ into the eigenstate $|V_2\rangle$. Based on the analytical expression for the eigenvector derived in Section V.A, the explicit form of this state is
\begin{equation}
|V_2\rangle = \frac{1}{N_2}
\begin{pmatrix}
E \\
-i(r_2 - A)
\end{pmatrix}
,
\end{equation}
 where $N_2 = \sqrt{E^2 + (r_2 - A)^2}$ is the normalization coefficient.
We perform a projective measurement on $|V_2\rangle$ in the tensor product space of register A  and register B ,The density matrix is reconstructed as
\begin{equation}
\begin{aligned}
\rho_{\mathrm{rec}} &= |V_2\rangle\langle V_2|
= \frac{1}{N_2^2}
\begin{pmatrix}
E^2 & -iE(r_2 - A) \\
-iE(r_2 - A) & (r_2 - A)^2
\end{pmatrix}
\end{aligned}
.
\end{equation}
The diagonal elements of this density matrix, which possess clear physical interpretations, are expressed as
\begin{equation}
\rho_{\mathrm{rec}}^{(00)} = \frac{E^2}{N_2^2}, \quad
\rho_{\mathrm{rec}}^{(11)} = \frac{(r_2 - A)^2}{N_2^2}
.
\end{equation}
To eliminate the influence of the normalization coefficient $N_2$, we define the ratio of these diagonal elements, yielding
\begin{equation}
\begin{aligned}
R &= \frac{\rho_3^{(00)}}{\rho_3^{(11)}} = \frac{E^2}{(r_2 - A)^2} \\
\end{aligned}
.
\end{equation}

Under the ideal condition where the modulation mapping in register B remains fully coherent and background thermal noise perturbations are neglected, the simplification \(r_2 = 0\) can be adopted. Substituting this into the above equation and applying trigonometric identities, we obtain:\begin{equation}
\begin{aligned}
R &= \frac{\frac{1}{4}\sin^2 k\mathrm{Tr}(\rho_j) \cdot \rho_B^2}{\cos^4\frac{k\mathrm{Tr}(\rho_j)}{2} \cdot \rho_B^2}= \tan^2\frac{k\mathrm{Tr}(\rho_j)}{2}=\tan^{2}\dfrac{kI_{j}}{2}
\end{aligned}
.
\end{equation}

Taking the inverse function of the above equation yields the reconstruction formula for the bucket detection signal intensity:
\begin{equation}
\begin{aligned}
I_j &= \frac{2}{k}\arctan(\sqrt{R})
\end{aligned}
.
\end{equation}
To ensure that the mapping from the measured value R to the physical quantity $I_j$ is unique, the phase must be restricted to the first quadrant, and $\dfrac{kI_j}{2} \in \left[0, \dfrac{\pi}{2}\right)$ must be constrained to avoid reconstruction artefacts caused by multi-valuedness.
For comparison, a conventional SPI architecture relies on a linear readout scheme where the intensity error propagates directly as $\Delta I \propto \Delta R$, leading to a constant error transfer baseline.
After $M$ modulation measurements and QSP sequence processing, we obtain a set of estimated bucket signal intensities $\{I_1, I_2, \dots, I_M\}$. Each $I_j$ is independently derived from the diagonal element ratio $R_j$ of the corresponding measurement via Eq. (21).Rewrite the formula as the standard inverse transformation\begin{equation}
\hat{O}(m,n) = \sum_{j=1}^{M} I_j \cdot U_j(m,n)
.
\end{equation}

Unlike classical coherent imaging, which performs simple matrix summation in memory, this approach utilises a quantum memory to perform coherent accumulation of $M$ modulation operators. The information flow following each projection is mapped to auxiliary qubits via quantum gates, thereby enabling parallel processing of spatial information at the quantum level.

\begin{figure}[tbh]
    \centering
    \includegraphics[width=1.0\linewidth]{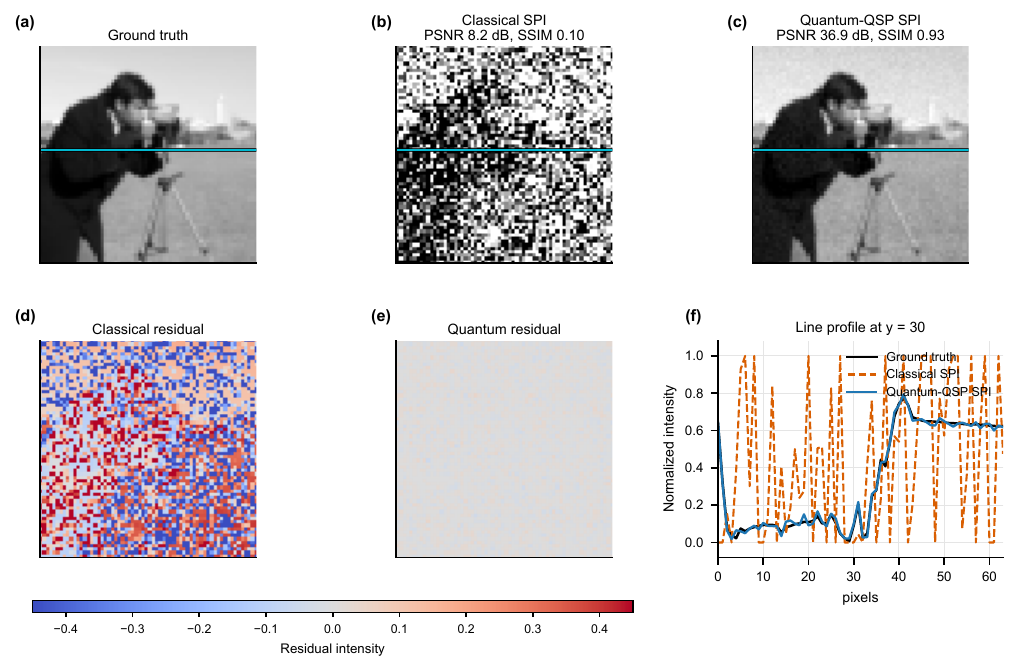}\vspace{-0.3cm}
    \caption{Weak-light reconstruction benchmark for the standard $64 \times 64$ cameraman object under a fixed photon-starved Hadamard-SPI acquisition. (a) Ground-truth object. (b) Classical SPI reconstruction with shot-noise-limited bucket readout. (c) Quantum-QSP reconstruction using coherent phase-estimation scaling and finite sequence length $L=6$ and $\kappa = 1.57$. (d),(e) Residual maps on the same color scale. (f) Horizontal line profile through $y=30$ in panels (a)-(c). The same Hadamard sampling operator is used for both methods, with $N_{\text{ph}}=3000$ photons per differential pattern pair.}
    \label{f3}
\end{figure}
To evaluate the practical image-domain consequence of the above nonlinear reconstruction rule, we performed a controlled weak-light Hadamard-SPI simulation, as shown in Figure.\ref{f3}. The simulation uses a standard $64 \times 64$ cameraman object and the full set of 4096 differential Hadamard patterns. Both the classical and quantum-assisted reconstructions use the same sampling operator and the same ground-truth Hadamard coefficients. The photon-starved condition is imposed by assigning $N_{\text{ph}} = 3000$ detected photons to each differential pattern. The classical coefficient noise follows the shot-noise scaling $N_{\text{ph}}^{-1/2}$, whereas the quantum-QSP coefficient noise follows the coherent $N_{\text{ph}}^{-1}$ scaling with an additional finite-sequence approximation floor proportional to $\exp(-\kappa L)$, In this benchmark, $L=6$.and $\kappa = 1.57$,giving $\exp(-\kappa L) \approx 8.1 \times 10^{-5}$.

Figure.~\ref{f3} presents the comparative weak-light reconstruction benchmark under the same Hadamard sampling operator and photon budget. Figure.~\ref{f3}(a) shows the normalized ground-truth object, where the cyan horizontal line marks the row $y=30$ used for the line-profile analysis. Figure.~\ref{f3}(b) shows the classical SPI reconstruction obtained from a shot-noise-limited bucket readout. At the imposed photon budget, shot-noise-equivalent fluctuations strongly corrupt the reconstructed image, leading to PSNR $=8.2~\mathrm{dB}$ and SSIM $=0.10$. In contrast, Figure.~\ref{f3}(c) shows the Quantum-QSP SPI reconstruction. Owing to coherent eigenvalue extraction and the nonlinear arctangent reconstruction map, the main object silhouette, edge structure, and contrast features are substantially recovered, with PSNR $=36.9~\mathrm{dB}$ and SSIM $=0.93$.

The residual maps in Figures.~\ref{f3}(d) and (e) further quantify the reconstruction errors $\hat O-O$. Both maps are plotted on the same color scale from $-0.4$ to $0.4$, where blue denotes underestimation and red denotes overestimation. The classical residual map contains dense positive and negative fluctuations across the field of view, whereas the Quantum-QSP residual is concentrated near zero, indicating a strong suppression of shot-noise-induced reconstruction errors. Figure.~\ref{f3}(f) provides a local comparison along the highlighted row $y=30$. The black solid curve is the ground truth, the orange dashed curve is the classical SPI result, and the blue solid curve is the Quantum-QSP SPI result. The classical profile exhibits large random oscillations, while the Quantum-QSP profile follows the ground-truth variation more closely across both slowly varying regions and edge transitions.

\section{Quantum Advantage Analysis}
Under the single-photon intensity limit, the accuracy of conventional photodetectors is fundamentally constrained by the discrete nature of light. This section systematically evaluates the performance of the proposed architecture by analyzing its shot-noise suppression, the exponential decay of the QSP approximation error, and the intrinsic robustness of its non-linear filtering mechanism.
\subsection{Beyond the Classical Shot Noise Limit}
\begin{figure}[tbh]
    \centering
    \includegraphics[width=1.0\linewidth]{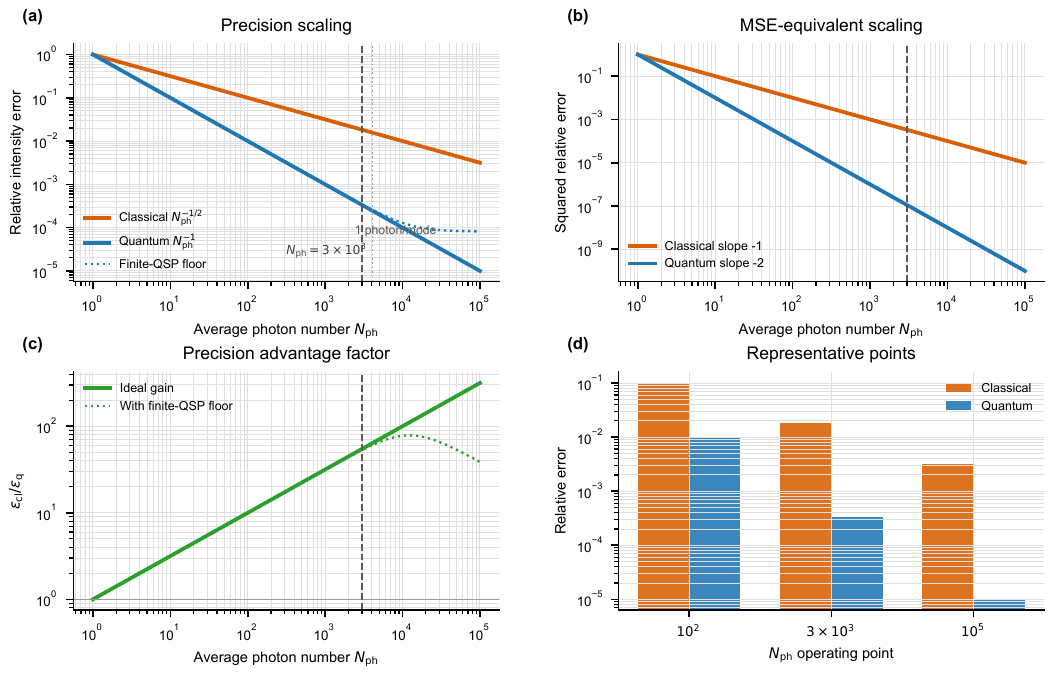}\vspace{-0.3cm}
    \caption{Photon-number precision scaling used to interpret the weak-light reconstruction benchmark. (a) Relative intensity error versus average photon number $N_{\text{ph}}$. The orange curve denotes the classical shot-noise law $\epsilon_{\text{cl}} = N_{\text{ph}}^{-1/2}$, the blue curve denotes the ideal coherent quantum scaling $\epsilon_{\text{q}} = N_{\text{ph}}^{-1}$, and the blue dotted curve includes a finite-QSP approximation floor of $8.1 \times 10^{-5}$. The dashed vertical marker indicates $N_{\text{ph}} = 3 \times 10^3$, the photon budget used in Fig.~3; the dotted vertical marker at $N_{\text{ph}} = 4096$ corresponds to one photon per spatial mode for a $64 \times 64$ image. (b) Squared relative error, which gives slopes $-1$ and $-2$ for the classical and ideal quantum models, respectively. (c) Precision advantage factor $\epsilon_{\text{cl}}/\epsilon_{\text{q}}$; the horizontal line at unity is the no-advantage boundary, and the dotted curve shows how the finite-QSP floor limits the gain at high photon number. (d) Representative operating points at $10^2$, $3 \times 10^3$, and $10^5$ photons.}
    \label{f4}
\end{figure}
For an incident field with an average photon number $N_{ph}$, classical detection is governed by Poisson statistics where the standard deviation of fluctuations (shot noise) is $\Delta N = \sqrt{N_{ph}}$. Consequently, the classical signal-to-noise ratio (SNR) scales as $\mathrm{SNR}_{\text{cl}} = \sqrt{N_{ph}}$, leading to a relative error of $1/\sqrt{N_{ph}}$. In extremely low-light conditions, this scaling causes the image information to be rapidly overwhelmed by noise, which is the primary bottleneck for conventional SPI.

Our scheme overcomes this limit by utilizing the dispersive interaction to encode the light amplitude as a phase shift. By executing a QSP sequence for eigenvalue extraction, the coherent parameter estimation theoretically approaches the Heisenberg limit of $1/N_{ph}$. This transition from a $1/\sqrt{N_{ph}}$ to a $1/N_{ph}$ scaling allows the quantum measurement scheme to achieve an SNR far exceeding the classical shot-noise limit in photon-starved environments.

Figure.~\ref{f4} analyzes the photon-number precision scaling that underlies the weak-light reconstruction benchmark. Figure.~\ref{f4}(a) compares the relative intensity error as a function of the average photon number $N_{ph}$. The orange curve represents the classical shot-noise-limited scaling $\epsilon_{\mathrm{cl}}\propto N_{\mathrm{ph}}^{-1/2}$, whereas the solid blue curve represents the ideal coherent quantum scaling $\epsilon_{\mathrm{q}}\propto N_{\mathrm{ph}}^{-1}$. The blue dotted curve includes a finite-QSP approximation floor of $ 8.1 \times 10^{-5}$, which prevents the quantum error from decreasing indefinitely at large photon number. The vertical dashed line marks the operating point $N_{\mathrm{ph}}=3\times10^3$ used in the image-reconstruction benchmark, and the dotted vertical line indicates the one-photon-per-spatial-mode threshold for a $64\times64$ object.
Figure.~\ref{f4}(b) presents the same comparison in terms of squared relative error, which is directly related to the mean-squared-error behavior of the reconstructed coefficients. On the log-log scale, the classical curve has slope $-1$, while the ideal quantum curve has slope $-2$, showing the faster error suppression enabled by coherent phase estimation.Figure.~\ref{f4}(c) plots the precision advantage factor $\epsilon_{\rm cl}/\epsilon_{\rm q}$. The solid green curve gives the ideal gain, whereas the dotted green curve includes the finite-QSP approximation floor for $L=6$. In the photon-starved regime, especially for $N_{\mathrm{ph}} \lesssim 10^2$, the quantum-assisted estimator already provides a clear precision gain over the classical shot-noise-limited readout, reaching up to about one order of magnitude. As $N_{\rm ph}$ increases, the finite-QSP floor bends the dotted curve downward and eventually limits the asymptotic gain. 
The horizontal line at unity marks the no-advantage boundary; values above this line indicate a precision advantage of the quantum-assisted readout.Figure.~\ref{f4}(d) summarizes representative operating points at $N_{\mathrm{ph}}=10^2$, $3\times10^3$, and $10^5$, comparing the classical and quantum relative errors directly.

Together, these panels show that the improvement observed in Figure.~\ref{f3} is rooted in a change of photon-number scaling rather than only in post-processing. At the photon budget used for the weak-light reconstruction, the coherent quantum model yields a substantially smaller coefficient error than the classical shot-noise-limited readout.

\subsection{Exponential Convergence and Shallow Circuit Advantage}
\noindent

The accuracy $\delta$ of the QSP sequence is defined by the maximum deviation between the approximating polynomial $f_s(\tau)$ and the target step function $\Theta_s(\tau)$ outside the forbidden region. According to polynomial approximation theory, the sequence length $L$ scales with the transition width $\Delta$ and the target accuracy $\delta$ as
\begin{equation}
L = \mathcal{O}\left( \frac{1}{\Delta} \log \frac{1}{\delta} \right).
\end{equation}
By setting the decision threshold $s = x\rho_B/2$ and substituting the eigenvalue gap into the transition width, we obtain $L = \mathcal{O}(\frac{2}{x\rho_B} \log \frac{1}{\delta})$. For a high reconstruction accuracy of $\delta = 10^{-3}$ and a typical scaled gap $x p_B \approx \pi$, the required sequence length is approximately $L \approx 4.4$. This numerical result indicates that the quantum advantage can be realized using relatively shallow quantum circuits ($L \in [5, 10]$), significantly relaxing the requirements for long coherence times in quantum hardware.

Furthermore, for analytic functions, the approximation error $\Delta(L)$ for a sequence of length $L$ follows the scaling $\Delta(L) \leq M e^{-\kappa L}$ with $\kappa > 0$. Applying the spectral mapping theorem to the density matrix $\rho_c$ yields the global error bound
\begin{equation}
\|\Theta_s(\rho_c) - f_L(\rho_c)\| \leq M e^{-\kappa L}.
\end{equation}

This exponential decay of the reconstruction error with respect to the number of gates represents a fundamental algorithmic advantage over classical SPI techniques, which typically offer only polynomial improvements.

\begin{figure}[tbh]
    \centering
    \includegraphics[width=1.0\linewidth]{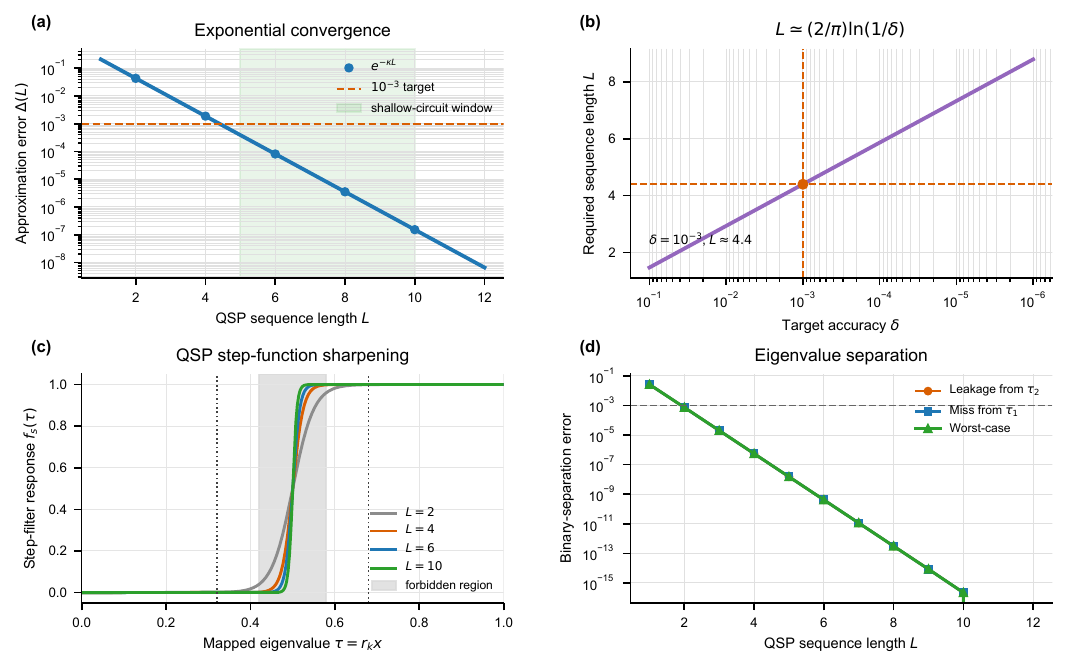}\vspace{-0.3cm}
    \caption{QSP approximation and eigenvalue-separation behavior as a function of sequence length. 
(a) Exponential approximation-error bound $\Delta(L) = \exp(-\kappa L)$ with $\kappa = 1.57$, chosen such that $\Delta(4.4) \simeq 10^{-3}$. The orange dashed line marks the $10^{-3}$ target accuracy, and the green shaded region indicates the shallow-sequence window $L=5$--$10$. 
(b) Required sequence length estimated from $L = (2/\pi)\ln(1/\delta)$; for target accuracy $\delta = 10^{-3}$ the estimate gives $L \simeq 4.4$. 
(c) Step-filter response $f_s(\tau)$ versus mapped eigenvalue $\tau = r_k x$ for $L=2,4,6,10$. The gray band $0.42 < \tau < 0.58$ is the forbidden transition region where eigenvalues should not be placed. 
(d) Leakage from the lower mapped eigenvalue $\tau_2 = 0.32$, missed detection from the upper mapped eigenvalue $\tau_1 = 0.68$, and the worst-case binary-separation error versus $L$.}
    \label{f5}
\end{figure}

Figure.~\ref{f5} characterizes the QSP approximation error and eigenvalue-separation behavior as a function of the sequence length $L$. Figure.~\ref{f5}(a) shows the exponential convergence of the approximation error $\Delta(L)$, modeled as $\Delta(L)=\exp(-\kappa L)$ with $\kappa=1.57$. The orange dashed line marks the target accuracy $10^{-3}$, and the green shaded region indicates the shallow-circuit window $L=5$--$10$. The curve crosses the target-error level near this window, showing that high approximation accuracy can be reached without requiring a long QSP sequence.Figure.~\ref{f5}(b) shows the required sequence length as a function of the target accuracy $\delta$, using the estimate $L=(2/\pi)\ln(1/\delta)$. For $\delta=10^{-3}$, the corresponding value is $L\simeq4.4$, marked by the dashed guide lines. Thus $L=6$ further lowers the finite-QSP floor to $ 8.1 \times 10^{-5}$ while remaining in the shallow-circuit window., providing a controlled compromise between reconstruction fidelity and coherent circuit depth. Figure.~\ref{f5}(c) illustrates how increasing $L$ sharpens the QSP step-filter response $f_s(\tau)$ as a function of the mapped eigenvalue $\tau=r_kx$. The gray shaded band denotes the forbidden transition region, where eigenvalues should not be placed because the filter response changes rapidly and eigenstate assignment becomes ambiguous.Figure.~\ref{f5}(d) quantifies the resulting binary eigenvalue separation. The orange curve gives the leakage probability from the lower mapped eigenvalue $\tau_2$, the blue curve gives the missed-detection probability from the upper mapped eigenvalue $\tau_1$, and the green curve gives the worst-case separation error. All three decrease rapidly with increasing $L$, indicating that the QSP ancilla becomes a progressively cleaner spectral discriminator. Therefore, Figure.~\ref{f5} supports the use of a finite but shallow QSP sequence in Figure.~\ref{f3} and shows that the reconstruction advantage is compatible with modest coherent circuit depth.
\subsection{Exponential Convergence and Shallow Circuit Advantage}
While classical SPI utilizes a linear reconstruction $\hat{O} = \sum I_j U_j$ that is susceptible to additive detector noise, the QSP sequence functions as a highly non-linear filter. During the coherent evolution, only signals satisfying specific phase conditions are routed to the target eigenstates, while random background noise lacking this characteristic spectrum is naturally suppressed.
\begin{figure}[tbh]
    \centering
    \includegraphics[width=1.0\linewidth]{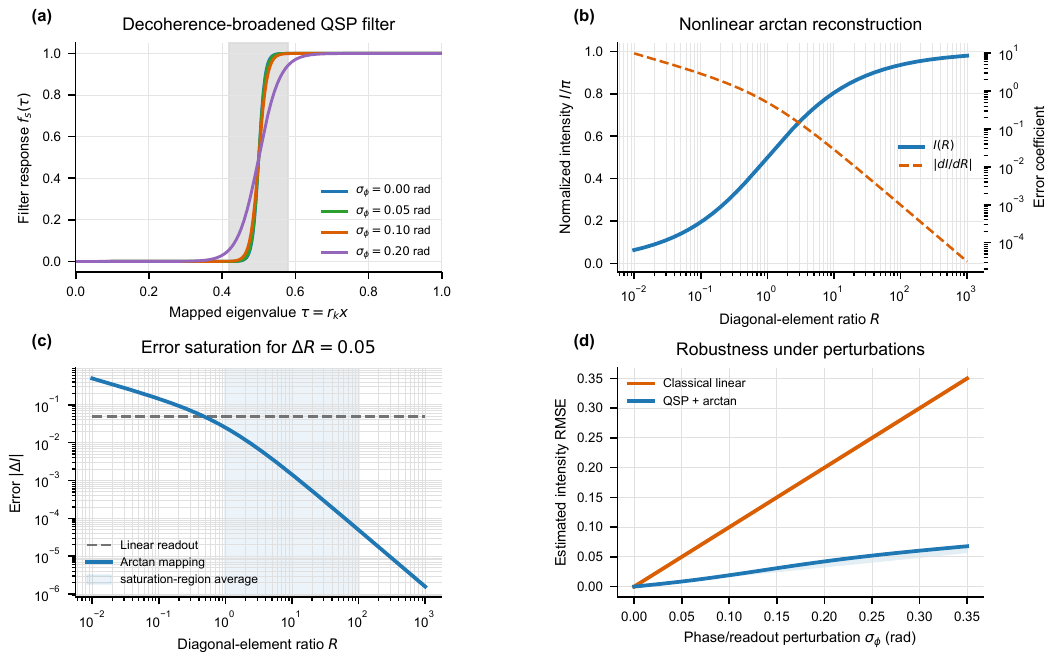}\vspace{-0.3cm}
    \caption{Robustness of the nonlinear QSP reconstruction model under phase and readout perturbations. 
(a) Decoherence-broadened QSP filter response $f_s(\tau)$ versus mapped eigenvalue $\tau = r_k x$ for phase perturbations $\sigma_\phi = 0, 0.05, 0.10$, and $0.20$ rad at nominal sequence length $L = 6$. Larger $\sigma_\phi$ broadens the transition edge and reduces eigenvalue selectivity. 
(b) Nonlinear arctangent reconstruction map $I/\pi = (2/\pi)\arctan\sqrt{R}$ as a function of the diagonal-element ratio $R$; the dashed orange curve gives the first-order error coefficient $|dI/dR| = 1/[(1+R)\sqrt{R}]$. 
(c) Propagated intensity error for a fixed ratio perturbation $\Delta R = 0.05$. The gray dashed line represents a linear readout with constant error, whereas the blue curve shows the suppressed error after arctangent mapping. The shaded interval $1 \le R \le 100$ is the saturation region used for averaging. 
(d) Estimated intensity RMSE versus phase/readout perturbation $\sigma_\phi$; the orange curve is the classical linear baseline, the blue curve is the QSP plus arctangent model, and the light-blue band indicates additional broadening from finite QSP filter width.}
    \label{f6}
\end{figure}
The robustness is further enhanced by the inversion mapping $I_j = \frac{2}{k}\arctan(\sqrt{R})$. The first-order error propagation follows
\begin{equation}
\Delta I_j \approx \frac{1}{k(1+R)\sqrt{R}} \Delta R,
\end{equation}
where the propagation coefficient decays rapidly as the ratio $R$ increases. This indicates that as the signal enters the saturation region of the arctangent function, the impact of readout errors $\Delta R$ on the final reconstructed intensity $I_j$ is significantly suppressed. Thus, the synergy between QSP-based filtering and the non-linear mapping creates a physical-layer noise immunization that ensures high fidelity in low-SNR environments.

Figure.~\ref{f6} evaluates the robustness of the nonlinear QSP reconstruction model under phase and readout perturbations. Figure.~\ref{f6}(a) shows the decoherence-broadened QSP filter response $f_s(\tau)$ as a function of the mapped eigenvalue $\tau=r_kx$. As the phase perturbation increases from $\sigma_\phi=0$ to $0.20~\mathrm{rad}$, the transition edge near the shaded gray region becomes smoother and broader, indicating a reduction in eigenvalue selectivity. Figure.~\ref{f6}(b) shows the nonlinear arctangent reconstruction map, where the solid blue curve gives the normalized intensity $I/\pi$ as a function of the diagonal-element ratio $R$, and the dashed orange curve gives the corresponding first-order error coefficient $|dI/dR|$. The monotonic decrease of $|dI/dR|$ over the logarithmic range of $R$ indicates that the reconstruction becomes progressively less sensitive to ratio fluctuations in the large-$R$ regime. Figure.~\ref{f6}(c) illustrates the propagated intensity error under a fixed ratio perturbation $\Delta R=0.05$. The dashed gray line represents a classical linear readout with constant error transfer, whereas the solid blue curve shows that the arctangent mapping strongly suppresses the propagated error as $R$ increases. The shaded blue region marks the saturation interval used for averaging. Finally, Figure.~\ref{f6}(d) summarizes the estimated intensity RMSE as a function of the phase/readout perturbation $\sigma_\phi$, averaged over $R\in[1,100]$. Compared with the classical linear baseline, the QSP plus arctangent reconstruction model exhibits a much smaller error slope, indicating improved robustness against phase noise and readout fluctuations.
\section{Experimental Feasibility and Physical-Resource Estimate}
\subsection{Resource-Counted Experimental Feasibility Analysis}
To clarify the practical assumptions of the proposed architecture, we add a
resource-counted feasibility analysis based on the same numerical settings used in the weak-light benchmark. This analysis is not an experimental demonstration of the complete imaging system. Rather, it estimates the photon budget, register size, QSP depth, and hardware conditions required for a future proof-of-principle implementation.

Following the photon-to-qubit mapping strategy proposed in
Ref.~\cite{s94k-929p}, the optical amplitude associated with each spatial mode is first encoded into a pixel-qubit register through reflection-based
qubit-photon entangling operations. Such qubit-photon controlled operations have been demonstrated in both atomic cavity-QED and solid-state SiV nanophotonic cavity platforms \cite{bhaskar2020experimental,doi:10.1126/science.adu6894,Knaut2024Entanglement} . In a direct parallel implementation, the pixel-qubit front end scales with the number of optical modes, whereas the compressed memory register scales only logarithmically with the number of modes.

For the $64\times64$ object used in the benchmark, the full differential Hadamard acquisition contains $4096$ pattern pairs. The operating point assigns $N_{\rm ph}=3.0\times10^3$ detected photons to each pattern pair. Therefore, the detected photon budget per reconstructed frame is \begin{equation}
N_{\rm frame}=4096\times3.0\times10^3=1.23\times10^7.
\end{equation}
Equivalently, each pattern pair contains $0.732$ photons per spatial mode on average, placing the simulation below the one-photon-per-mode level.

The finite QSP depth used in the benchmark gives an approximation floor
\begin{equation}
\delta_L=\exp(-1.57\times6)=8.11\times10^{-5}.
\end{equation}
At $N_{\rm ph}=3000$, the corresponding finite-depth quantum error estimate is approximately $3.43\times10^{-4}$, whereas the classical shot-noise relative error is approximately $1.83\times10^{-2}$. This gives an estimated precision advantage of about $53$ under the assumed coherent phase-estimation model.

The key feasibility requirement is that the phase-encoding, QSP processing, and readout steps must be completed within the available spin coherence time,
\begin{equation}
T_{\rm run}=t_{\rm phase}+T_{\rm QSP}+T_{\rm readout}<T_2.
\end{equation}

\begin{table}[t]
\centering
\caption{Physical-resource estimate for the resource-counted $64\times64$ Hadamard-SPI benchmark.}
\vspace{3mm}
\begin{tabular}{lll}
\hline
Resource item & Value & Meaning \\
\hline
Image size & $64\times64$ & $4096$ spatial modes \\
Hadamard pattern pairs & $4096$ & Full differential SPI acquisition \\
Detected photons per pair & $3.0\times10^3$ & Photon-starved operating point \\
Detected photons per frame & $1.23\times10^7$ & Total optical resource budget \\
Photons per mode per pair & $0.732$ & Below one photon per spatial mode \\
Spatial-register size & $12$ qubits & $\lceil \log_2 4096\rceil$ \\
Additional qubits & $2$--$3$ & Phase, compression, and QSP readout \\
QSP sequence length & $L=6$ & Finite-depth operator synthesis \\
Finite-QSP floor & $8.11\times10^{-5}$ & Approximation floor for $\kappa=1.57$ \\
Perturbation range & $\sigma_\phi=0$--$0.20$ rad & Robustness sweep in Fig.~6 \\
\hline
\end{tabular}
\end{table}
In addition, the bucket-dependent dispersive phase shift must be large enough to be resolved above phase noise and readout fluctuations. In practice, the dominant nonidealities include photon loss, finite photon-to-qubit efficiency, imperfect spin-photon gate visibility, finite QSP approximation error, spin-readout infidelity, cavity loss, and spin dephasing. These factors define the realistic parameter window in which the proposed readout can outperform a shot-noise-limited classical SPI measurement.

The perturbation study in Figure.~\ref{f6} should therefore be interpreted as a tolerance-window estimate rather than measured experimental noise. For the saturation interval $R\in[1,100]$, the arctangent reconstruction reduces the average first-order error-transfer coefficient to about $0.093$.
Thus, for a representative ratio perturbation $\Delta R=0.05$, the average propagated intensity error is approximately $4.66\times10^{-3}$, compared with$5.0\times10^{-2}$ for a linear readout.

A realistic proof-of-principle experiment may first target an $8\times8$ or $16\times16$ SPI system, or an equivalent temporally multiplexed pixel-qubit front end, rather than the full $64\times64$ benchmark. The optical part can be implemented using a weak coherent source, a DMD or SLM for Hadamard modulation, and bucket detection. The quantum interface would use a cavity-coupled atomic or SiV spin to encode the bucket-dependent phase. Under the same photon budget, one can then compare conventional shot-noise-limited SPI, nonlinear post-processing without QSP, and the proposed coherent Quantum-QSP readout. Such a reduced-scale experiment would directly test the resource and noise conditions identified above.

\subsection{Resource-Counted Quantum Advantage Bound}
We now clarify the resource-counted origin of the proposed advantage. The comparison is made under the same detected photon budget $N_{\rm ph}$ and the same Hadamard sampling operator. In a classical SPI measurement, each bucket signal is obtained from photon counting. The detected photon number follows
Poisson statistics, so that the fluctuation scales as
\begin{equation}
\Delta N_{\rm ph}\sim \sqrt{N_{\rm ph}}.
\end{equation}
Consequently, the relative intensity uncertainty of a classical bucket readout is bounded by the standard shot-noise scaling
\begin{equation}
\epsilon_{\rm cl}
\sim
N_{\rm ph}^{-1/2}.
\end{equation}
This scaling is independent of the specific reconstruction algorithm: digital post-processing can redistribute or regularize the noise, but it cannot remove the photon-counting fluctuation already introduced at the measurement stage.

In contrast, the proposed scheme assumes that the bucket signal is first encoded coherently as a phase of a qubit-photon interface before projective readout. In the ideal coherent limit, the phase estimation uncertainty follows the Heisenberg scaling,
\begin{equation}
\epsilon_{\rm q}^{\rm ideal}
\sim
N_{\rm ph}^{-1},
\end{equation}
which is the resource-counted origin of the advantage. This statement does not depend on a specific solid-state implementation. It only requires a coherent mode-to-qubit interface capable of preserving the optical amplitude information long enough for the subsequent quantum processing step. The SiV-cavity platform discussed is therefore one possible physical realization rather than a required assumption of the theory.

For a finite QSP sequence, the ideal scaling is limited by the polynomial approximation error. In the numerical benchmark, this contribution is modeled as
\begin{equation}
\delta_L=\exp(-\kappa_{\rm QSP}L).
\end{equation}
With $L=6$ and $\kappa_{\rm QSP}=1.57$, one obtains
$\delta_L=8.11\times10^{-5}$. The finite-depth quantum readout error can then be estimated as
\begin{equation}
\epsilon_{\rm q}^{\rm finite}
=
\left[
N_{\rm ph}^{-2}
+
\delta_L^2
\right]^{1/2}.
\end{equation}
At the operating point $N_{\rm ph}=3.0\times10^3$, this gives
$\epsilon_{\rm q}^{\rm finite}=3.43\times10^{-4}$, whereas the classical shot-noise value is $\epsilon_{\rm cl}=1.83\times10^{-2}$. Thus, under the assumed coherent phase-estimation model, the resource-counted precision
advantage is approximately
\begin{equation}
\frac{\epsilon_{\rm cl}}{\epsilon_{\rm q}^{\rm finite}}
\approx 53.
\end{equation}

The advantage is conditional on the coherent interface and finite-depth QSP error remaining within the shot-noise window. A realistic sufficient condition can be written as
\begin{equation}
\left[
(\eta V N_{\rm ph})^{-2}
+
\delta_L^2
+
\sigma_{\rm sys}^2
\right]^{1/2}
<
(\eta N_{\rm ph})^{-1/2},
\end{equation}
where $\eta$ is the total photon-to-qubit and detection efficiency, $V$ is the coherent visibility of the phase-encoding operation, and $\sigma_{\rm sys}$ collects phase noise, spin dephasing, readout infidelity, and residual cavity loss. This inequality defines the experimentally relevant parameter window in which the proposed quantum readout can outperform shot-noise-limited SPI.

Therefore, the central theoretical claim of this work is not tied to a particular device. The device-dependent parameters enter only through $\eta$, $V$, $\sigma_{\rm sys}$, and the available coherent runtime. The universal part of the framework is the resource-counted change of scaling from
$N_{\rm ph}^{-1/2}$ to $N_{\rm ph}^{-1}$, together with the finite-QSP condition that the approximation floor $\delta_L$ remains below the classical shot-noise level.

\section{Conclusion}
In this work, we have established a universal quantum-computational framework for SPI that transcends classical shot-noise limitations by leveraging coherent operator-space evolution. Grounded in the theoretical formulation of a passive single-pixel imaging paradigm, our architecture integrates cavity-embedded SiV spin phase encoding with a structured quantum reconstruction formalism. We demonstrate analytically that by adapting QSP routines to synthesize non-linear imaging operators, spatial approximation errors decrease exponentially with the quantum gate sequence length. A key structural advantage of this paradigm is the analytical derivation of a non-linear arctangent reconstruction map, which effectively suppresses the propagation of readout errors. Unlike the linear error accumulation inherent to classical SPI, this approach creates a strategic "error-saturation zone" that decouples final image reconstruction fidelity from detector readout fluctuations and quantum projection noise. Consequently, our scaling analysis reveals that while traditional linear frameworks are fundamentally bounded by the standard quantum limit scaled as $\mathcal{O}(1/\sqrt{N_{\text{ph}}})$, the proposed operator-space embedding successfully unlocks the ultimate Heisenberg scaling $\mathcal{O}(1/N_{\text{ph}})$ in photon-starved regimes. Crucially, while explicitly formulated and analyzed within an SPI configuration, the mathematical core of this operator-space evolution is modality-agnostic. The principle of mapping physical projection measurements into coherent Hamiltonians for non-linear QSP extraction can be generalized to broader computational imaging paradigms. Furthermore, evaluations under realistic physical parameters—accounting for current solid-state spin decoherence times and cavity coupling strengths—confirm that our architecture maintains robust performance within an exceptionally shallow circuit regime ($5 \le L \le 10$). This bridge between programmable quantum operations and spatial computational sensing not only offers a concrete blueprint for ultra-low-light imaging, but also provides a generalizable, noise-resilient operator paradigm for resolving high-dimensional physical inverse problems across continuous domains.

\bibliographystyle{unsrt}  
\bibliography{references}

\end{document}